# Rate-Induced Transitions in Networked Complex Adaptive Systems:

Exploring Dynamics and Management Implications Across Ecological, Social, and Socioecological Systems

Vítor V. Vasconcelos[✝1,2,3], Flávia M.D. Marquitti[✝4,5], Theresa Ong[✝6], Lisa C. McManus[✝7], Marcus Aguiar[4], Amanda B. Campos[8], Partha S. Dutta[9], Kristen Jovanelly[6], Victoria Junquera[10], Jude Kong[11], Elisabeth H. Krueger[12], Simon A. Levin[10,13], Wenying Liao[14], Mingzhen Lu[15], Dhruv Mittal[1], Mercedes Pascual[16], Flávio L. Pinheiro[17], Juan Rocha[18], Fernando P. Santos[1], Peter Sloot[1,2], Chenyang (Crispy) Su[6], Benton Taylor[14], Eden Tekwa[22], Sjoerd Terpstra[19,20], Andrew R. Tilman[21], James R. Watson[24], Luojun Yang[13], Senay Yitbarek[23], Qi Zhan[16]

1. Informatics Institute, University of Amsterdam, Amsterdam, The Netherlands
2. Institute for Advanced Study, University of Amsterdam, Amsterdam, The Netherlands
3. Center for Urban Mental Health, University of Amsterdam, Amsterdam, The Netherlands
4. Instituto de Física Gleb Wataghin, Universidade Estadual de Campinas, Brazil
5. Instituto de Biologia, Universidade Estadual de Campinas, Campinas, Brazil
6. Department of Environmental Studies, Dartmouth College, USA
7. Hawai'i Institute of Marine Biology, University of Hawai'i at Mānoa, Kāne'ohe, HI, 96744, USA
8. Programa de Pós Graduação em Ecologia, Universidade de São Paulo, São Paulo, Brazil
9. Department of Mathematics, Indian Institute of Technology Ropar, India
10. High Meadows Environmental Institute, Princeton University, USA
11. Department of Mathematics and Statistics, York University, Canada
12. Institute for Biodiversity and Ecosystem Dynamics, University of Amsterdam, Netherlands
13. Department of Ecology and Evolutionary Biology, Princeton University, USA
14. Department of Organismic and Evolutionary Biology, Harvard University, USA
15. Santa Fe Institute, USA
16. Department of Ecology and Evolution, University of Chicago, USA
17. NOVA IMS - Universidade Nova de Lisboa, Portugal
18. Stockholm Resilience Center, Stockholm University, Sweden
19. Institute for Marine and Atmospheric research Utrecht, Utrecht University, Netherlands
20. Graduate School of Informatics, University of Amsterdam, Netherlands
21. Northern Research Station, USDA Forest Service, USA
22. Department of Biology, McGill University, Canada
23. Department of Biology, The University of North Carolina at Chapel Hill, USA
24. College of Earth, Ocean and Atmospheric Sciences, Oregon State University, USA

[✝]Lead authors, corresponding authors

**Author contributions:**
VVV, FMDM, TO, and LCM conceptualized and performed the research. All authors contributed to writing and editing the manuscript. VVV, FMDM, TO, and LCM wrote the first and final drafts, and all authors reviewed the final draft prior to submission.

# Abstract

Complex adaptive systems (CASs), from ecosystems to economies, are open systems and inherently dependent on external conditions. While a system can transition from one state to another based on the magnitude of change in external conditions, the rate of change—irrespective of magnitude—may also lead to system state changes due to a phenomenon known as a rate-induced transition (RIT). This study presents a novel framework that captures RITs in CASs through a local model and a network extension where each node contributes to the structural adaptability of others. Our findings reveal how RITs occur at a critical environmental change rate, with lower-degree nodes tipping first due to fewer connections and reduced adaptive capacity. High-degree nodes tip later as their adaptability sources (lower-degree nodes) collapse. This pattern persists across various network structures. Our study calls for an extended perspective when managing CASs, emphasizing the need to focus not only on thresholds of external conditions but also the rate at which those conditions change, particularly in the context of the collapse of surrounding systems that contribute to the focal system's resilience. Our analytical method opens a path to designing management policies that mitigate RIT impacts and enhance resilience in ecological, social, and socioecological systems. These policies could include controlling environmental change rates, fostering system adaptability, implementing adaptive management strategies, and building capacity and knowledge exchange. Our study contributes to the understanding of RIT dynamics and informs effective management strategies for complex adaptive systems in the face of rapid environmental change.

# Introduction

We are embedded in complex adaptive systems (CASs)—e.g., ecosystems (Dasgupta, 2021; Levin, 1999), social networks (Centola, 2021), financial markets (May et al., 2008), or food-production networks (Coe et al., 2008)—and we depend on the products and services they provide. CASs can undergo abrupt and largely irreversible changes known as regime shifts, which may present opportunities (Mark Meldrum et al., 2023; Nyborg et al., 2016) or challenges (Pecl et al., 2017; Wernberg et al., 2016). Traditional approaches to understanding regime shifts focus on identifying critical environmental thresholds using steady-state analyses such as, in the context of the climate crisis, a 2°C limit to the average planetary temperature (Lenton et al., 2019; Staal et al., 2020). However, these approaches assume stationarity of the environment, while recent theories have identified rate-induced transitions (RITs) as another important phenomenon that can occur in dynamical systems (Ashwin et al., 2012; Kaur and Dutta, 2022; Neijnens et al., 2021; Siteur et al., 2016; Wieczorek et al., 2010). RITs occur when a sufficiently high rate of change in external conditions—not (just) the magnitude of change—causes a regime shift. Here, we argue that RITs may be a common feature of CASs and occur when external environmental, technological, or social conditions change at a rate faster than the system's ability to adapt. As such, these RITs have significant implications for the stability and resilience of all CASs.

CASs often comprise interconnected networks—be it food chains, social media platforms, urban living, or production and supply chains—where each node within these networks can be subjected to rapid, substantial changes, which may trigger cascading regime shifts (Rocha et al., 2018). Such cascades, similar to domino effects, may emerge when external changes like climate fluctuations, propagation of misinformation, or global migration patterns occur at unprecedented rates. Amplified by network topology, these changes ripple across the system, influencing the internal rates and pattern of change within the network. For example, the flow of (mis)information diffusion on social platforms are influenced by user interconnections (Vosoughi et al., 2018) and the interaction between noise and network dynamics (Czaplicka et al., 2013). Environmental changes can induce and cut off migration patterns that ripple across global networks (Aguirre and Tabor, 2008). Within ecosystems, the interdependence of species can be formulated, e.g., as plant-pollinator networks and their overall resilience to environmental changes (Bascompte and Scheffer, 2023). The complex dynamics of these networks significantly impact internal rates of change, potentially sparking cascade effects and creating conditions for faster transmission and diffusion. Thus, understanding network structures and their adaptability to external changes becomes crucial in predicting and managing rate-induced regime shifts.

Decades of theory development have resulted in a better understanding of how CASs can pass tipping points and undergo regime shifts, from collapsing ecosystems like forests and coral reefs to the degradation of urban livelihoods due to rapid urban growth (Boulton et al., 2022, 2013; McManus et al., 2020; Scheffer et al., 2001). However, recent events such as the 2008 financial crisis, Brexit in 2016, Californian and Australian wildfire seasons over the past years, the Covid-19 pandemic, or the need for rapid sociotechnological change suggest two outstanding problems: 1) large-scale regime shifts are happening more frequently and 2) we are still greatly underskilled at anticipating or controlling these kinds of events. The urgency of recognizing RITs lies in the challenge of scale and pace mismatches and is essential for enhancing the resilience of our ecological, social, and governance systems.

The primary component of an RIT is an external element (i.e., an element not affected by the system) that is changing and that can change at different rates. Many of the world's processes are increasing their pace. In the natural world, increasing emissions of $CO_2$ drive ecological change at all scales (Rama et al., 2022). Socially, internet access expands and disrupts ways of communicating and living (Castells, 2011) and urbanicity affects our mental health (Wal et al., 2021). Furthermore, the development and deployment of technology accelerate with population and economic growth, information flows increase with the automation of data analysis, and data-driven models of world markets lead to near-instantaneous transactions (Bettencourt et al., 2007; Brynjolfsson and McAfee, 2014; Castells, 2011).

Adaptive responses by system components at lower levels of organization may allow systems to respond and adapt to change and perturbation, but there is no guarantee that these will increase systems resilience. Indeed, species regrow, accurate information can be learned, and new firms are created, but the emergent consequences are unpredictable (Anderson, 1972).

In general, when external conditions change slowly, systems start to settle into a new equilibrium. This often means that the internal structure of the CAS changes. The new equilibrium can be a state that is similar to the previous one or the system may transition from one basin of attraction—states of the system that share the same future—to another; slow-time-scale variation may change the dynamic topology qualitatively, eliminating basins (Strogatz, 2018; Thom, 1969). In some cases, especially in managed systems, modification of internal structure may provide resilience (Levin et al., 2022; Slobodkin and Rapoport, 1974). Some species adapt and migrate as temperatures and resource availability change, preventing extinction under changing environments (Pecl et al., 2017). Some schools now teach students the perils of online misinformation (Department of Education, 2023), and there is a growing effort to highlight and prevent the spread of false information by creating new tools to attain factual and contextualized information (Lazer et al., 2018). Most financial systems recently developed capital buffers to prevent economic bubbles and market crashes to prevent system-wide issues (Haldane and May, 2011). In other cases, however, the result is system collapse (Centeno et al., 2023).

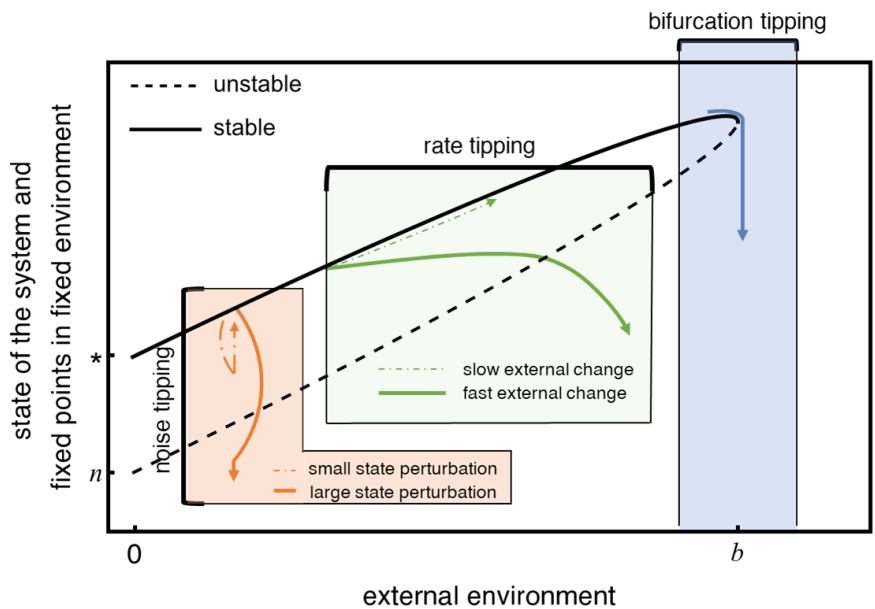

**Figure 1.** Different types of transitions away from a stable regime. The stable regime, represented by the solid black line, is considered in the absence of change in the external environment as well as in the absence of perturbation. When the system is below the orange line, it moves towards a (not represented) different regime. Orange represents a *noise-induced transition*. There, a small enough perturbation from the stable line allows the system to recover back to it (dashed orange line). However, if the perturbation goes beyond some threshold value, represented as a dashed line, the system moves away to a different regime (full orange line). The area that allows for recovery is called the basin of attraction of the stable regime. In blue, *bifurcation-induced tipping* concerns the dynamics of the system around a critical value of the external environment, $b$. Below $b$, the stable regime exists. As the external environment changes, the stable regime can also change until a point when it suddenly disappears, leading to a runaway of the system into a different regime. Reverting the external environmental conditions does not return the system to the initial regime, indicating hysteresis. Finally, in green, we have *rate-induced transitions*. These transitions also concern external environmental changes. However, the stable regime exists for all values of the external environment. Thus, any small change in external environments cannot lead to a regime shift. However, if the speed of the external environmental change is high enough, the system is unable to track the quasistatic attractor (stable regime) and a regime shift occurs.

CASs exhibit regimes that are dynamic and multidimensional, represented by topological dynamics that track slow-variable, possibly exogenous changes (Thom 1969). We can start to classify these behaviors by using simple one-dimensional models that also exhibit a variety of transition behaviors. Figure 1 illustrates three different types of transitions away from a stable regime (full black line) in a single (one-dimensional) system plotted for different external environments (i.e., an element that affects the system but is not affected by it). In orange, we depict noise-induced transitions, which happen when random fluctuations or noise in a system cause it to switch between different stable states or regimes. These transitions are not driven by a change in the external environment but rather by the inherent randomness or noise within the system. In blue, we represent bifurcation-induced transitions. This type of transition occurs when a system's behavior changes dramatically due to a (small) variation in the external environment, leading the system to a new regime. Finally, in green, a rate-induced transition occurs when a system experiences a transition due to the speed at which the external environment changes, rather than due to a specific value of the external environment itself. In this type of transition, the system may not have enough time to adapt to the changing conditions, leading to a different regime.

# Examples of RITs across systems

In this section, we explore examples of rate-induced transitions across various systems, including ecological, social, and socio-ecological systems. We aim to demonstrate the relevance of RITs in different contexts and highlight their importance in understanding and managing interconnected CASs. The next section is dedicated to the network aspect of these CASs.

## Ecological systems

Ecological systems, or ecosystems, are complex networks of living organisms and their physical environment that interact as cohesive units. They encompass the flow of energy and nutrients between various biotic and abiotic components, maintaining a balance that supports diverse life forms. Ecological systems, such as coral reefs, forests, and phytoplankton systems, are often considered prototypical complex adaptive systems. Beyond their interconnected elements, they are characterized by their collective capacity to adapt to changes in their environment. Diversity with natural selection, rapid evolution, gene transfer, adaptation, and plasticity are all examples of mechanisms that allow ecological systems to deal with external change while maintaining ecosystem functioning. When external change outpaces these adaptive mechanisms, it may lead to rate-induced tipping, while limited or absent adaptive mechanisms lead to bifurcation-induced tipping. Rapid environmental changes in ecological systems include climate change, habitat fragmentation, and pollution. Box 1 illustrates three ecological systems—coral reefs, forests, and phytoplankton systems—and how they might suffer from rate-induced transitions.

## Social systems

Social systems can be divided into many subdomains. Socio-technological systems refer to the interplay between society and technology, examining how technological advancements influence social structures and human behavior. Socio-political systems describe the interactions between social groups and political institutions, shaping governance, power dynamics, and policy-making. Socioeconomic systems encompass the relationship between social and economic factors, exploring how social inequalities, wealth distribution, and resource allocation affect society's overall well-being. Individuals, organizations, cities, and countries must constantly develop to meet changing demands and environments. In that sense, the ecological shifts as well as technological, political, and economic changes and developments present themselves as external stressors faced by society, where specific elements include climate change, disruptive technologies, the spread of false information, conflict, human mobility, economic shifts, and rising inequality. Fortunately, these social systems are characterized by high innovation or creativity rates, contextual and intergenerational learning, redistribution, and reorganization: mechanisms that allow them to adapt to novel circumstances without major disturbances to the structures that support them. While creativity might be unbounded, the rate at which they produce and implement solutions to growing problems may be outpaced, generating rate-induced transitions. Box 1 illustrates specific examples of rate-induced transitions in socio-technological systems (e.g., disruptive technological innovation and demographic changes), socio-political systems (e.g., Brexit and accelerated opinion dynamics), and socio-economic systems (e.g., sudden economic crises or financial market crashes).

## Socioecological systems

A socioecological system is an integrated framework that examines the complex interactions between human societies and their surrounding natural environments. This approach recognizes the interdependence of social and ecological processes (Levin et al., 1998) and aims to promote sustainable resource management, conservation efforts, and inclusive wealth (Arrow et al., 2014; Dasgupta and Levin, 2023). Classic examples include fisheries, forestry, agricultural systems, and water management, where rules and institutional structures adapt to resource availability and balance the risk of collapse with high service levels, and, in turn, resources are directly and indirectly affected by human use (Ostrom, 2015). The most paradigmatic case of a socioecological system is the stewardship of the Earth (Chapin et al., 2010). Changing our governing structures and economies at a fast enough rate will be crucial to face climate change and maintain civilization and its structures as we know them. RIT can happen as disruptive events in these paced changes, but they can also occur for promoting the necessary rapid switch from the status quo into sustainable practices.

| | |
|---|---|
| Ecological Systems | **Coral reefs** face threats from increasing coral bleaching caused by marine heatwaves (Hughes et al., 2018). Coral bleaching occurs when the symbiosis between the coral animal and its photosynthetic dinoflagellate is disrupted, resulting in symbiont expulsion and exposed coral skeletons (e.g., (Glynn, 1996; Rädecker et al., 2021)). Identifying coral adaptation mechanisms is crucial for reef protection and recovery. One adaptation pathway involves evolving higher optimal growth temperatures (Dixon et al., 2015). Modeling studies suggest that coral populations exhibit sensitivity to the pace of temperature increase, not only to the magnitude of temperature change (Gil et al., 2020). Slow temperature increase may enable local selection and gene flow to facilitate higher optimal growth temperature evolution. However, rapidly increasing temperatures can lead to declines in individual reefs and reef networks (McManus et al., 2020). Coral bleaching thresholds are influenced by photosymbiont thermotolerance (Berkelmans and van Oppen, 2006; Buerger et al., 2020; Swain et al., 2017), and coral adaptation can occur through photosymbiont evolution towards increased thermotolerance (Buerger et al., 2020; Császár et al., 2010) or symbiont shuffling towards more thermotolerant clades (Berkelmans and van Oppen, 2006). In some locations, corals may exhibit alternative stable states between coral- and macroalgal-dominated states (Mumby et al., 2007). Rate-induced tipping in coral could |
| | interact with bifurcation-induced tipping in coral-algal systems through stressor interaction, potentially leading to unanticipated tipping into a coral depauperate state (Gil et al., 2020). |
| | **Forests** are a prime example of a complex adaptive system (Levin, 1998), composed of multiple subsystems (e.g., the soil bacterial community and fungal networks, plant communities, and resident animals) that are linked to each other by processes such as plant-microbe interactions (Lu and Hedin, 2019; Smith and Read, 2008), pollination (Bascompte et al., 2003; Waser and Ollerton, 2006), and seed dispersal (Bascompte et al., 2003; Levey et al., 2002). This networked structure means that for a given rate of environmental change, various subsystems within a forest will respond to change at dramatically different rates. For instance, rate differences between soil and air temperature changes seem to cause zombie fire reemergence in peatlands (O'Sullivan et al., 2022). Generally, the microbial community can track rapidly rising temperature through rapid change in community composition (timescale of hours to days (Bardgett and Caruso, 2020)), while the response rate of vascular plants is fundamentally constrained by their slow growth, recruitment and species turnover (timescale of decades to centuries (Stephenson and van Mantgem, 2005)). While rate-induced transitions have never been explicitly tested in forests, forest dieback due to rapidly intensifying droughts (e.g., (Anderegg et al., 2012)) presents a useful framework for these ideas. Had temperatures and drought intensity increased slowly over centuries, tree communities would likely have been able to gradually shift toward more drought-resistant species. However, the rapid rise in drought intensity seen in recent decades has created full-scale forest dieback as resident trees are unable to survive in drier conditions and more drought-resistant tree species cannot immigrate via dispersal quickly enough. Whether these landscapes are re-colonized by drought-resistant trees, shift to grass-dominated systems, or take on some other ecosystem state will likely depend on how potential mismatches in the response rates of the microbial, plant, and disperser communities impact the connections between them. On the other hand, noise by itself in bistable vegetation soil can induce stability (D'Odorico et al., 2005), but the combination of noise and rate of change remains unexplored in ecological systems. |
| | **Phytoplankton** are primary producers and the basis of energy flow and material circulation of the whole aquatic ecosystem. Their growth depends on two essential resources: nutrients and light (Zhang et al., 2021). Rate-induced tipping points in phytoplankton populations can occur due to changes in temperature (Vanselow et al., 2022), but also potentially due to changes in light intensity via alterations in the biomass of other phytoplankton groups. As light intensity increases, photosynthesis induces growth and, in turn, shading. If light intensity increases too fast, the shading of phytoplankton becomes insufficient to prevent photoinhibition, leading to decreased photosynthesis rates. The rate of changing light intensity can trigger a tipping point in phytoplankton populations and have cascading effects on the aquatic ecosystem. |
| Social Systems | **Sociotechnological and economic dynamics** As technologies evolve and new opportunities emerge, economic actors, such as cities and businesses, are pressured to adopt new technologies and business models, thereby improving the efficiency of their processes. Currently, with the increase in job automation, changing market needs requires these actors to adapt their knowledge base from manual to cognitive skills. Failing to do so can risk being outcompeted, where shifting economic patterns can make cities, regions or countries that have fallen behind less attractive for businesses, resulting in economic stagnation, high levels of unemployment and inequality, or even bankruptcy (Frank et al., 2018). Furthermore, human migration, driven by economic-, climate-, or conflict-driven reasons requires adequate adaptation of organizations, cities, and countries to maintain functional, stable societies. These developments are path-dependent processes, influenced by existing knowledge-related structures, termed the Knowledge Space (Alabdulkareem et al., 2018; Alstott et al., 2017; Guevara et al., 2016; Hartmann et al., 2021; Hidalgo, 2021; Hidalgo et al., 2007; Pinheiro et al., 2022). The literature highlights a tendency |

for development towards more closely related activities, which implies a slow pace of adaptation, and can lead economies into development traps that exacerbate disparities between agents (Hartmann et al., 2020). Diversification of unrelated knowledge and economic structures, although rare, is essential for structural transformations (Pinheiro et al., 2022), as it allows more agile adaptation to changing demands. Therefore, economic actors face the challenge of identifying optimal development paths to keep evading structural "traps" and promote inclusive development through adaptation at rates that match technological advancements, and allow responding to pressure resulting from technological advancements and competition.

**Political systems and opinion dynamics** The political governance system, which encompasses decision-making processes, is coupled with the system of opinion dynamics of the masses. However, there is an inherent difference in the timescales of this coupled dynamical system. Bureaucratic processes and the reshaping of political systems are generally slow when compared to the dynamics of social networks, particularly online media like Twitter and Facebook (Gros, 2017). This difference in timescales can also be attributed to the decentralization of these systems. Government organizations have a time delay as they are elected after a fixed number of years. On the other hand, cascades of news, opinion expression, and misinformation on social media networks are further accelerated due to both the content and the algorithms designed to maximize user engagement. If policies track public opinion, an accelerated rate of opinion dynamics may trigger a rate-induced transition in the political system. Enhancing rates of social integration in a context of global change and migration is essential in such processes. Such transitions could provide insights into the emergence of polarization and the rise in global instability (Kuo, 2019). Understanding the interactions between political governance and opinion dynamics, focusing on their differing timescales and decentralized nature, is crucial for examining the stability and adaptability of democracies in the face of rapidly evolving public opinion.

**Urban growth** The majority of the global population now lives in cities, and its fraction is expected to increase to 68% by 2050 (United Nations, Department of Economic and Social Affairs, 2019). This growth is caused by structural socioeconomic changes, as well as climate- and conflict-driven migration. Although larger cities are associated with increased economic productivity (Lobo et al., 2013), they also exhibit inequality in access to services and income (Brelsford et al., 2017; Pandey et al., 2022) and relate to higher prevalence of common mental disorders (Guyot et al., 2023; Wal et al., 2021). Sustainable urban growth requires infrastructure expansion for resources and services; however, many cities, particularly in the Global South, struggle to accommodate this growth, leading to the development of informal settlements and potential increase in segregation. Gradual infrastructure degradation and increasing frequency and multiplicity of shocks, such as climate impacts and land use change, can cause rate-induced transitions in urban systems when the rate of maintenance and repair is exceeded by the rate of degradation (Klammler et al., 2018; Krueger et al., 2022, 2019). Urban water infrastructure operation and maintenance costs often exceed revenues, and urban growth or more frequent extreme events can exacerbate these costs and hinder adaptation efforts. Urban systems' adaptability to change relies on state/municipal-level funding and management efficacy and individual or household-level economic and social capital. As management-level adaptability becomes overwhelmed, the need for adaptation shifts to individuals, but the two levels remain interdependent.

**Technological innovation and regulation/governance dynamics** The recent speed up in innovation in artificial intelligence and applications (Tang et al., 2020) (e.g., ChatGPT/OpenAI) has triggered a debate around the possible positive impacts of AI in many areas of society, such as Education, Health, Science, Transportation, Economy, and Environmental Sustainability. However, there are also negative impacts of AI that require the design of new governance

| | |
|---|---|
| | frameworks to mitigate labor displacement, inequality, credit score initiatives, and the negative consequences of an AI race (Dafoe, 2018; Rampersad, 2020). Combined with the speed of AI development that exceeds the pace of governance and regulatory institutions and the natural adaptation of socio-technological and economic sectors by many orders of magnitude, the rate-induced transition that advanced AI systems could be responsible for dramatic transformations where the adverse outcomes are significantly greater than the reaped benefits. |
| | **Financial systems** The collapse of financial institutions is often followed by a cascade of failures (dictated by the network of financial institutional interdependence) that can have drastic social and economic consequences (Elliott et al., 2014, p. 2; Haldane and May, 2011; May et al., 2008). Although market regulators put in place mechanisms and conditions to ensure the resilience of financial institutions (controlling the institutional systemic risk), these do not take into consideration the possible impact of the rate of change of the environment (e.g., central bank reference interest rates, house market bubbles) and the institutional ability to adapt to such changes (e.g., portfolio adjustments, and exposure ), leading to unexpected events that can have negative implications for society. |
| Socioecological Systems | **Socioecological systems** have varying abilities to adapt to externally changing environments. These systems are dynamic and complex, with human societies (ideally) adjusting their behavior or practices to accommodate new ecological conditions, while ecosystems may respond by regenerating lost species or altering their structure in the presence of human use (Deb et al., 2023). Both co-adapt to larger-scale external changes. Namely, human perceptions can change depending on how we measure and respond to the environment and as a function of the quality of our measurements. However, there are limits to these adaptation processes, especially in the context of rate-induced transitions, where rapid changes can push these interconnected systems beyond their capacity to adjust and maintain stability.<br>In some cases, a rapidly declining resource growth rate may cause a system to lag behind its steady state, leading to higher consumption than predicted by steady-state analysis. Beyond a critical rate of change, this can result in overconsumption and depletion of the resource (Siteur et al., 2016). Socioecological systems that involve harvesting resources, such as fisheries, are also subject to adaptation limits. Institutional adaptation rates, which determine how quickly harvesting strategies change in response to ecological and economic factors, can play a crucial role in these systems (Tekwa and Junquera, 2022; Tilman et al., forthcoming). When institutions are slow to adapt, systems can exhibit alternative stable states, with resource levels either remaining high or becoming unsustainable depending on initial conditions (Tekwa et al., 2019). Overall, the ability of socioecological systems to adapt to changing environments varies, and limits to adaptation can be observed in the context of rate-induced transitions. These limits can be influenced by factors such as institutional adaptation rates and the presence of noise, which can ultimately determine the sustainability of resource use and the resilience of the system. |

**Box 1.** Examples of Rate-Induced Transitions in Ecological, Social, and Socioecological Systems and their properties

## A Local Model for RITs

Well-known bifurcation forms and noise-induced critical transitions (see Figure 1) hinge on the idea of escaping from attractors, configurations that reflect a stable but potentially dynamic regime of a system. However, adaptability challenges concepts of stability in such systems. Changing a control parameter is typically visualized as a shifting position of a static attractor that represents the dynamics of the system's properties. However, in an adaptive system, there is a reshaping of the attractor itself such that it ultimately reverts to its original or equivalent form, something traditionally viewed as structural adaptation.

In this section, we study RITs in CASs through a local model of a single CAS. Later, we extend the model to a network where elements comprise this (sub)system. The model captures various types of critical transitions by encompassing recovery mechanisms, structural adaptation, and external rates of change, offering a comprehensive understanding of critical transitions across different CASs.

The model focuses on a CAS characterized by a potentially multidimensional state, encapsulated in $x$, representing all elements that contain feedback among them (see Figure 2). The CAS is affected by external variables, $E$, which we refer to as the system's external environment. The external environment influences the state of the system, $x$, through its distance from the (changing) ideal external environmental condition, $\epsilon$. The CAS can adapt to changes in $E$ by changing $\epsilon$, but $E$ is not affected by it, by definition of the external environment.

The CAS can sustain shocks to its state and recover but may tip into a different state if shocks are large enough, resulting in an irreversible change of state (noise-induced transition). Further, the relationship between the state of the system, $x$, and the external environment, $E$, is described by the normal form (Strogatz, 2018) of a saddle-node bifurcation when $b[E-\epsilon] = 1$, guaranteeing the existence of bifurcation-induced transitions for static $\epsilon$:

$$\frac{dx}{dt} = -x^2 + b[E - \epsilon].$$

The function $b[E-\epsilon]$ represents the effects of the environment on the system and is built such that the system presents the highest resilience when the external conditions, $E$, coincide with the internally defined ideal conditions, $\epsilon$. The simplest formulation of such a dependence is

$$b[E - \epsilon] = 1 - (E - \epsilon)^2.$$

Structural adaptation of the CAS is incorporated into the model by allowing the CAS to change its ideal external environment, $\epsilon$ (a variable endogenous to the CAS), to match the current external environment, $E$, at a rate, $\alpha$:

$$\frac{d\epsilon}{dt} = \alpha(E - \epsilon).$$

Allowing for, e.g., a dependence of $\alpha$ in $E$ that brings $\alpha$ to zero creates b-tipping at a specific value of the external environment. For simplicity we assume $\alpha$ is fixed for now and let it depend on adjacent CAS in the next section. Combining these, we obtain a non-autonomous system that captures the dynamics of $x$ as the environment changes at a fixed rate, $r$ (see full derivation in SI):

$$\frac{dx}{dt} = -x^2 + 1 - \left(\frac{r}{\alpha}(e^{-\alpha(t-t_0)} - 1) + (E[t_0] - \epsilon[t_0])e^{-\alpha(t-t_0)}\right)^2.$$

With this model, we can analyze the behavior of the system under two distinct phases of the external environment (see inset Figure 3): Phase I, where the environment—starting at an arbitrary value $E=0$—changes at a fixed rate, $r$, and Phase II, where the environment stabilizes (rate of change equal to zero) at a new value, $E_{max}$. The total environmental change is then given by $E_{max}$ and the rate of change in Phase I by $r$. (Terpstra et al., 2023) Combining the phases, the model helps identify the different regimes of the CAS and the conditions under which the system can sustain the stable regime, always undergo irreversible tipping, or exhibit a combination of these behaviors depending on the rate of environmental change (RIT).

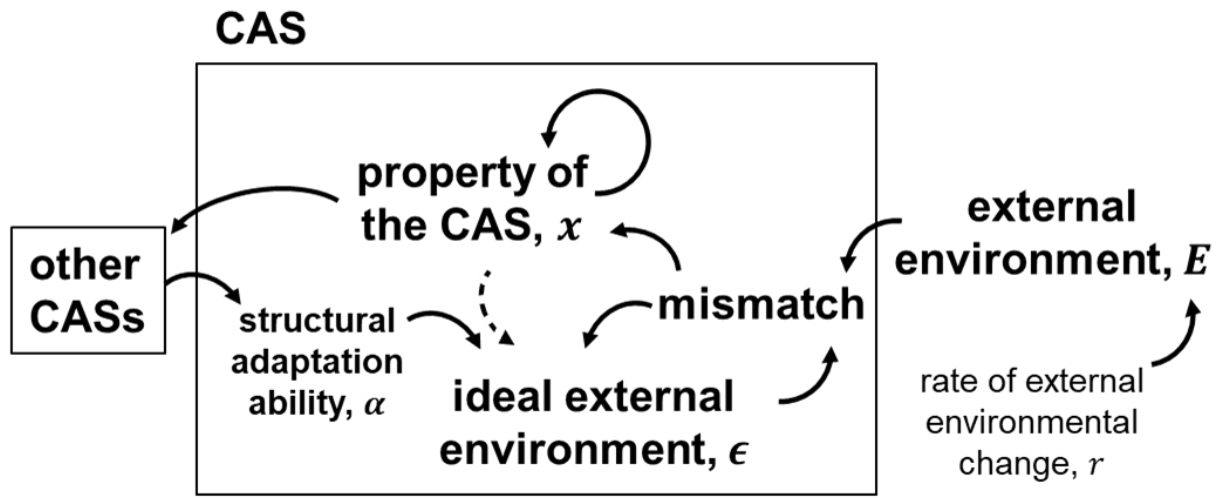

**Figure 2.** Visual representation of the model, showcasing its main components and how they interact. The CAS represents the components of the system that contain feedback, in this case, a property of interest of the CAS, $x$, and a level of match with the environment. This match mediates the dynamics of $x$, and is conceptualized through an ideal condition, $\epsilon$, that can adjust through structural adaptation ability, $\alpha$. The property of interest, $x$, can also mediate the effect of $\alpha$ in changing $\epsilon$ (dashed line). This last mediation is not present in the model we consider here. The properties of other CASs may affect structural adaptation ability, creating a network of CASs that we explore in the next section.

The model accounts for different types of critical transitions, including noise-, bifurcation-, and rate-induced transitions and their combination. The resilience of the system is reflected by its ability to (partially) recover the property of interest for low enough perturbations in a constant environment, the ability to structurally adapt to new environmental conditions to fully or partially recover that property of interest, and the capacity to withstand a variety of parameter changes (Holling, 1973) and, the contribution of this work, a variety of parameter rates of change.

By incorporating these diverse aspects, the model provides a comprehensive understanding of resilience in CASs and offers insights into the dynamics and management implications across ecological, social, and socio-ecological systems. Namely, the model highlights the importance of classifying boundaries for i) the minimum value for the property of interest (avoiding noise-induced transitions), ii) the state of the external environment (avoiding bifurcation-induced transitions), and iii) the rate of change of the external conditions (avoiding rate-induce transitions). Ignoring the latter is equivalent to assuming an infinite ability to structurally adapt as long as there is a critical mass for the property of interest, which might lead to unanticipated regime shifts.

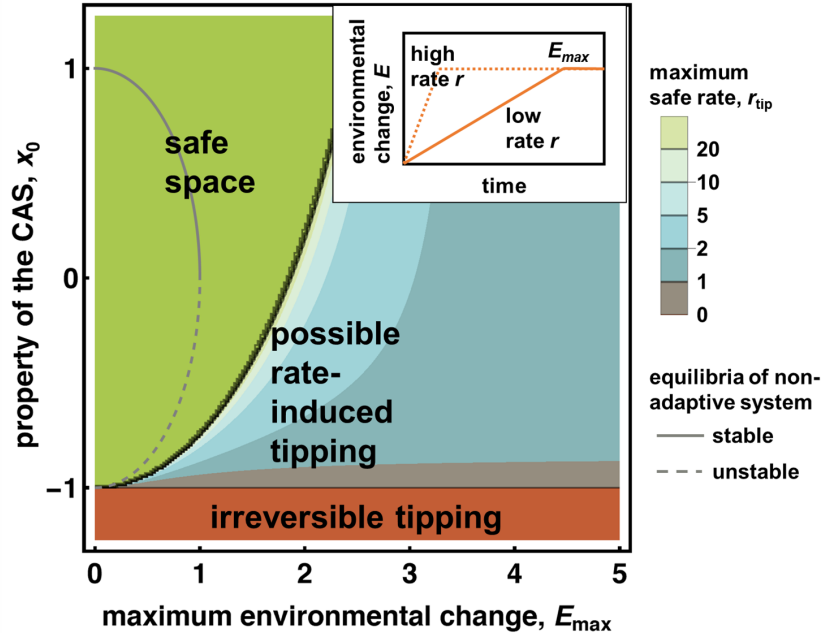

**Figure 3**: Regimes of complex adaptive systems derived from the local model. A CAS going through the external environmental change highlighted in the inset: the external environment changes at a given rate, $r$, up to a maximum value, $E_{max}$. Given $E_{max}$, the rate of external environmental change $r$, and the starting conditions of the property of interest of the system $x_0$, the CAS is characterized as being in either a 'safe space' (green), where it always tracks a prosperous state, here normalized at 1, or it can always undergo 'irreversible tipping' into a different state (orange-brown) and degrade beyond -1. A third region exists where either can happen (shades of blue). In this third region, it is not only the value of maximum environmental change and the initial state of the system that determines whether the system is safe but also the rate at which environmental change happens (maximum safe rate, $r_{tip}$). A structural adaptation ability of $\alpha = 1$ is considered. When the CAS has no structural adaptation ability ($\alpha = 0$), the 'safe space' is the region above the unstable equilibrium delineated by the gray dashed line, and the 'irreversible tipping' is the complementary region. For extremely adaptive systems, the safe-space region takes over the region where possible rate-induced tipping can exist and only noise-induced transitions can occur.

# RITs in Networked Systems

Complex systems typically consist of many interacting elements, each with adaptive capacities to withstand shocks and adapt to external conditions. Adaptive capacity often derives from interactions with other subsystems, interconnected by processes such as diffusion, communication, migration, nutrient flows, infrastructure, and (online and offline) social and information networks. Thus, the adaptive rate of a system can depend on its connectivity in a network with other systems.

## Networks as sources of adaptive capacity of CASs

In the context of corals, networks significantly influence the adaptive potential of coral reefs under increasing sea surface temperatures. Coral-symbiont networks, which operate on a local scale, can indicate coral vulnerability to bleaching based on the thermal optima of associated symbionts (Berkelmans and van Oppen, 2006; Swain et al., 2017; Williams and Patterson, 2020). Generalist corals, which can interact with multiple symbiont clades, may increase the overall adaptive response to rising temperatures in a coral reef (Mies et al., 2020). Furthermore, dispersal networks connect coral reef patches in a meta-community, potentially facilitating the repopulation of degraded reefs by warm-adapted corals (Matz et al., 2020; McManus et al., 2021a). The dispersal of photosymbionts, such as the invasive zooxanthellae from the Indo-Pacific, can impact coral thermal vulnerability and growth rates (Pettay and LaJeunesse, 2013). The structure of these networks, particularly the properties and states of hubs, influences the rate of adaptation to changing temperatures (Hagen et al., 2012; McManus et al., 2021b, 2021a). Understanding the structure of coral dispersal networks for coral larvae and photosymbionts across the world's reefs can provide valuable insights into their vulnerability to increasing temperatures and overall adaptive potential.

Networks also play a crucial role in the structural adaptation and resilience of forest ecosystems, which are hierarchically networked complex adaptive systems (Levin 1998). These ecosystems consist of multiple tiers of subsystems, such as soil bacteria, fungi, plants, and animals, that interact through processes like plant-symbiont interactions (Smith and Read, 2008), pollination (Potts et al., 2010), and seed dispersal (Janssen et al., 2006). Forest subsystems respond to environmental changes like rising temperatures (Zeppel et al., 2013), nitrogen deposition (Aber et al., 1989; Vitousek et al., 1997), or reduced precipitation (Betts et al., 2004) at varying rates, with microbial communities adapting more quickly than vascular plants (Bardgett and Caruso, 2020; Stephenson and van Mantgem, 2005)(Stephenson et al. 2005). Networked rate-induced transitions may be present in temperature- and drought-induced forest dieback via differing adaptation rates (Anderegg et al., 2013, 2012). Rapid environmental changes can lead to system-level regime shifts and potentially increased reliance of vascular plants on their symbiotic partners for adaptation. As the most rapidly adapting subsystem, microbes influence soil resource availability, soil structure, and pathogen prevalence, which in turn determine plant community composition (Phillips et al. 2013). Thus, the interconnectedness of forest subsystems through networks is essential for their resilience and adaptation to changing environments.

The topology of urban infrastructure networks, including roads, traffic, water and greenery distribution, and sewer systems, affects their vulnerability and resilience (Kalapala et al., 2006; Wal et al., 2021; Yazdani and Jeffrey, 2011; Zhan et al., 2017), and the functionality of these networks is crucial for urban economies and human well-being. There is an increasing and convergent complexity and interdependence of urban systems (Klinkhamer et al., 2019), which can change their overall resilience (Chen et al., 2015; Gao et al., 2014). As these networks mature, they typically evolve into truncated power-law topologies, making them more resilient to random failures but more vulnerable to targeted attacks on hub nodes (Krueger et al., 2017; Yang et al., 2017). Rapid urban growth in the Global South challenges the careful planning and implementation of resilient urban infrastructure networks. Historically, urban infrastructure degradation has been linked to urban collapse, as seen in the ancient city of Angkor (Penny et al., 2018). The structure and modularity of these networks can help cities repair damages and adapt flexibly to rapid growth (Yazdani and Jeffrey, 2011). Modular approaches can also facilitate experimentation and identify feasible solutions to keep pace with the transformative forces driving urban dynamics.

In political systems, the decision-making process is coupled with the dynamics of public opinion. However, there is an inherent difference in the time scales of these coupled systems (Axelrod, 1997; Liggett, 1999). Bureaucratic processes and reshaping of political systems are slow, particularly when compared to the rapid dynamics of social networks and online media like Twitter and Facebook (Castells, 2015). The difference in time scales can be attributed to the degree of decentralization and the fixed intervals between elections (Tarrow, 2011). The spread of news, opinions, and misinformation on social media networks is accelerated by both content and algorithms designed to reshape interaction networks and maximize user engagement (Allcott and Gentzkow, 2017; Bakshy et al., 2015). If policies aim to track public opinion, then this accelerated rate of opinion dynamics may trigger rate-induced transitions in the political system (Centola, 2018). Such transitions could provide insights into the emergence of polarization (Sunstein, 2018) and consequences for socially beneficial coordination (Vasconcelos et al., 2021) and the rise of instability around the world, as social networks and social media platforms play an increasingly influential role in shaping opinions and driving political change (Bennett and Segerberg, 2012).

## A networked model of RITs

We extend the local model that we proposed above for a single CAS by considering a collection of CASs, each as an individual element or node of an interconnected larger system. Each CAS of this larger system contributes to the structural adaptability of other elements. We use complex networks to describe the interactions between the different CASs considered such that the overall system is a larger CAS. These interactions are considered as the contribution of the (state or) property of interest of each CAS to the structural adaptability of a focal CAS (see left side of Figure 2). Thus, the contribution of node $j$ to the structural adaptive capacity of node $i$ can be encapsulated in an interaction matrix, $a_{ij}$, and the structural adaptation parameter for node $i$ can now be written as

$$\alpha_i = \alpha_i^* \frac{1}{<k>} \Sigma_j a_{ij} \max(0, x_j + 1)/2,$$

where $<k> = \frac{1}{N}\Sigma_{i,j} a_{ij}$ is the average degree of the network. This formulation asserts that, in equilibrium under a constant environment, the average structural adaptability is just the average $\alpha_i^*$, while guaranteeing that nodes that have tipped (with $x_i < -1$) do not contribute to the structural adaptability of other nodes. The results in Figure 4 show that these coupled CASs exhibit similar properties as the original single CAS, with rate-induced transitions occurring at a critical value of the rate of external change, $r$. Different networks (and network structures) exhibit different tipping points, but the pattern is maintained. Overall, low-degree nodes tip first since they have fewer connections and, thus, less adaptive capacity. High-degree nodes tip even though they have more connections because the source of their adaptability (lower-degree nodes) has previously collapsed.

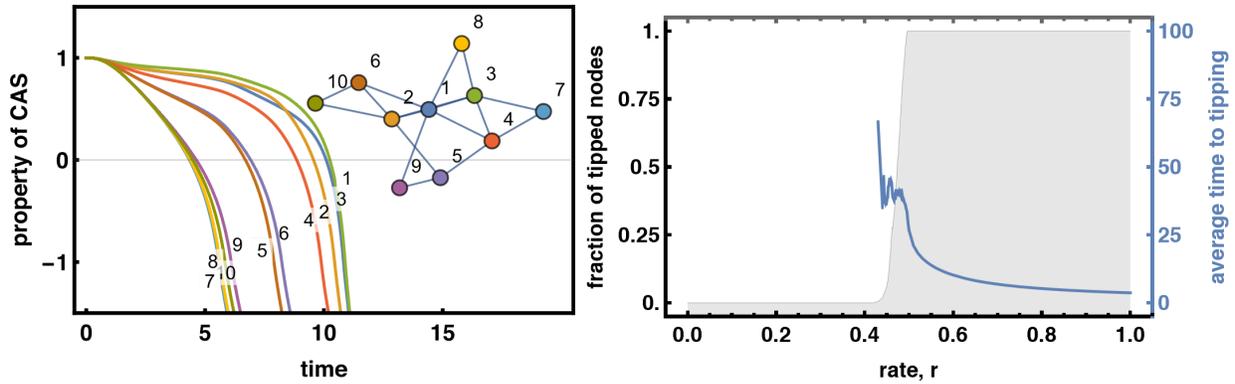

**Figure 4: Rate-induced transitions in networked structures.** (Left) An illustrative example of a small network with 10 nodes, with an external rate of change of r = 0.7, and average structural adaptability $\alpha_i^* = 1$. As time increases, nodes cannot adapt and thus they tip, reducing the adaptive capacity of other nodes, which contributes to the tipping of other nodes. Nodes are ranked from high to low degree, and nodes with higher degree tip later (see also appendix, Figure S3, for the larger networks used in the right panel) as their effective degree is reduced. (Right) Different rates of change in the environment alter the outcome of 30 Barabasi-Albert (BA) networks of 1000 nodes, where 2 new nodes are added with each iteration of the BA algorithm (see SI for more homogeneous networks). For rates of external change bigger than a critical value, all nodes in the networks go through rate-induced tipping. Below the critical value, there is a small range of rate values where some networks are able to resist collapse while others collapse.

# Management Implications

Understanding and addressing rate-induced transitions (RITs) in complex adaptive systems (CASs) has significant implications for managing ecological, social, and socio-ecological systems. In light of the findings presented, we discuss potential strategies and approaches to mitigate the impacts of RITs and enhance the resilience of these systems.

**Identify the threats**

*Monitor and control the rate of environmental change*: One of the key insights from our study is the importance of the rate at which external conditions change. The results suggest managers and policymakers may benefit from closely monitoring and, where possible, regulating the pace of environmental change to keep it within the adaptive capacity of the system when the transition is undesirable. For example, in ecological systems, this may involve controlling not just the extent but also the rate of habitat fragmentation or deforestation; in socio-technological systems, it could involve implementing regulations to manage the rapid spread of technology or information (Lazer et al., 2018; "Pause Giant AI Experiments," 2023; Quinn and Milmo, 2023). Guaranteeing stable external environments (social and ecological) can help each of the systems (ecological and social). Importantly, our understanding and prediction of system responses requires considering the 'slowness' of a system, or the delay in the system's response to environmental changes. This delay can often obscure the system's true state (Faassen et al., 2015). On the contrary, where transitions are desirable, rate of implementation may be crucial for system change.

*Identify mismatches between the rates of social and environmental process and governance processes*: The temporal (and spatial) scale at which natural resource governance takes place frequently does not match the scales at which social and environmental dynamics unfold (Levin et al., 2013; Meyfroidt et al., 2022). Thus, beyond monitoring the rate of change of social and ecological processes, paying attention to the possible mismatch between the rate at which governance and management interventions (can) occur and the rate at which natural resource dynamics unfold may also benefit policymakers.

*Develop early warning indicators*: Implementing robust monitoring systems targeted to detect early warning signs that include impending RITs can help managers and policymakers take proactive measures to prevent or mitigate their impacts. These early warning indicators need to become more mechanistic, precise for particular systems, and go beyond analysing streaming of data for properties derived theoretically considering the systems are quasistationary and close to equilibrium. They may include shifts in key system parameters, the emergence of new feedback loops, or changes in the distribution of resources within the system.

**Develop adaptive action plans**
*Foster adaptability and structural adaptation*: Enhancing the adaptive capacity of systems can help mitigate the risks of RITs, increasing their ability to cope with rapid environmental change. For ecological systems, this may involve conserving and enhancing biodiversity and genetic variation. For sociotechnological and socioeconomic systems, this may involve encouraging innovation, learning, and structural adaptation. In the context of networked CASs, identification of nodes with outsized influence (for example, with high out-degree and, thus, influencing the adaptability of many other subsystems) can assist in prioritization for conservation or investment in order to leverage network-level adaptive capacity (Pires et al., 2017) and prevent or enhance indirect effects of networked systems overall (Pires et al., 2020).

*Measure outcomes and test RIT explanations against alternatives*: Regime shifts are widespread across social and ecological systems, but evidence of these transitions has not been universally accepted in the scientific community. Part of the problem is that data on both environmental change and system outcomes have been absent or collected in only a few instances of a class of systems proposed to share similar behaviors. Another problem is the lack of general statistical methods to estimate parameters in models featuring abrupt transitions and to evaluate explanatory power. Policymakers can take actions today to mitigate the potential existence of RITs, but precise policies are likely to require investments in monitoring and analyzing system responses.

*Implement adaptive management strategies*: Adopting adaptive management approaches that incorporate continuous learning, experimentation, and adjustment could help systems better cope with the uncertainties and complexities associated with RITs. This may involve testing different interventions or policies to determine their effectiveness in promoting resilience and adjusting them as new information becomes available.

**Foster collaborations**

*Build capacity and knowledge exchange*: Strengthening the capacity of individuals, organizations, and communities to understand and respond to RITs is essential for promoting resilience in complex systems. This may involve investing in education, research, and training programs, as well as fostering collaboration and knowledge exchange among diverse stakeholders. Understanding networked RITs highlights the need for cross-sector collaboration in early-warning systems, such that when one system fails, in the relatively narrow window of opportunity, others can rapidly increase preparedness to avoid cascading collapse across linked systems.

# Conclusion

This paper presents a novel model that captures rate-induced transitions (RITs) in complex adaptive systems, emphasizing the importance of the rate of environmental change and the adaptive capacity of the system, including in the presence of multiple complex adaptive subsystems interrelated via complex networks. The relevance and applicability of RITs are demonstrated across various systems, including ecological, social, and socioecological systems, showcasing their significance in understanding and managing these systems effectively. The identification of key properties of RITs, as well as the factors contributing to their occurrence, provides valuable insights for designing management strategies that promote resilience and adaptability in the face of rapid environmental change.

The management implications derived from the study's findings can guide decision-makers in mitigating the impacts of RITs and enhancing the resilience of complex systems. This can be achieved by controlling the rate of environmental change, fostering adaptability, and implementing adaptive management strategies. This paper highlights the importance of knowledge creation and capacity building in promoting resilience and mitigating the impacts of RITs across ecological, social, and socioecological systems.

Future research directions and potential applications of this work include expanding the model to consider additional factors and feedback mechanisms, applying the model to specific case studies, investigating the effectiveness of various management strategies in preventing or mitigating RITs, and developing tools and frameworks to support decision-makers. Namely, the model is specified such that the nodes contribute to each other's adaptability. However, in specific circumstances, more connectivity could also mean transmitting shocks faster and imply that nodes would tip faster if they are more connected. By studying RITs in different systems and leveraging the insights gained from this work, we can advance our understanding of the dynamics and consequences of these transitions, ultimately informing more effective and resilient management strategies for complex adaptive systems.

# Acknowledgments

The authors acknowledge BIRS for providing the necessary conditions for the fruitful discussions and stimulating environment that occurred during the 'BIRS workshop [22w5067]: Rate-Induced Transitions in Networked Systems,' which made this article possible. VVV and DM acknowledge funding from ENLENS through the project "The Cost of Large-Scale Transitions: Introducing Effective Targeted Incentives." FMDM acknowledges the Coordenação de Aperfeiçoamento de Pessoal de Nível Superior – Brazil (Finance Code 001). M.L. would like to acknowledge the support by the Santa Fe Institute Omidyar Fellowship. ABC thanks Coordination of Superior Level Staff Improvement - CAPES Brazil (grant number 88887.630562/2021-00) and São Paulo Research Foundation - FAPESP Brazil (grant number 2022/06847-4). The findings and conclusions in this publication are those of the authors and should not be construed to represent any official USDA or U.S. Government determination or policy.

# Supplementary Information —

Rate-Induced Transitions in Networked Complex Adaptive Systems: Exploring Dynamics and Management Implications Across Ecological, Social, and Socioecological Systems

## Model details

CAS are known for being able to sustain shocks to their state. However, if shocks are large enough, they may tip into a different state, i.e., go through an irreversible change of state. Consider the following canonical system showing irreversibility, the normal form of a saddle-node bifurcation[1], where $b$ is represented as a function of $E - \epsilon$,

$$\frac{dx}{dt} = -x^2 + b[E - \epsilon], \qquad (1)$$

where we arbitrarily set $b[0] = 1$. Under ideal conditions, $\epsilon = E$, this system has a normalized stable configuration at $x = 1$. When in this configuration, this CAS can sustain shocks that change its internal state and return to equilibrium, as long as the shock does not bring $x$ below (a normalized value of ) -1. The range of states, from -1 to 1 (and above 1) is the basin of attraction of a prosperous regime (e.g., a state where biomass, abundance, productivity is high), $x = 1$. When the shock brings the system outside that basin, the system will degrade, going through a noise-induced transition. As the environment changes, $b$ varies and the prosperous state's basin of attraction is also modified. The basin is defined by the union of $\left(-\sqrt{b}, \sqrt{b}\right)$ and $\left(\sqrt{b}, \infty\right)$, when the prosperous stable state at $x = \sqrt{b}$ exists. If $b$ goes below zero, the system loses its stable state and, irrespectively of the state, it will degrade. In those conditions, the basin of attraction of the prosperous state disappears and the system is said to go through a bifurcation-induced transition.

Often, CASs can respond not only to perturbations to its internal state, $x$, but they can also adapt to perturbations to the environment by adapting to a new environment. Since $\epsilon$ is the ideal external environmental condition for the system, $b$ is the highest when $E$ coincides with $\epsilon$. Thus, we write

$$b[E - \epsilon] = 1 - (E - \epsilon)^2. \qquad (2)$$

For simplicity, and to explicitly control the structural adaptive capacity of this CAS, we allow the system to change its ideal environment, $\epsilon$, to match the current environmental state, $E$, at a fixed rate, $\alpha$, independent of the state of the system, resulting in

$$\frac{d\epsilon}{dt} = \alpha(E - \epsilon) \qquad (3)$$

For a fixed environmental state $E$, it is trivial to prove that the same two solutions described above ($x = \sqrt{b}$, stable, and $x = -\sqrt{b}$, unstable) exist with $\epsilon = E$. As the environment changes, $E[t]$, the ideal environmental condition also changes as

$$\epsilon[t] = e^{-\alpha t}\left(\epsilon[0] + \alpha \int_0^t e^{\alpha \tau} E[\tau] d\tau\right).$$

When the environment changes at a fixed rate, $r$, i.e., $E[t] = E[0] + rt$,

$$\epsilon[t] = \frac{r}{\alpha}\left(e^{-\alpha t} - 1\right) + (E[0] - \epsilon[0])e^{-\alpha t} + E[0] + rt.$$

Plugging these in Eq.(2) and, then, in Eq.(1), we get the non-autonomous system

$$\frac{dx}{dt} = -x^2 + 1 - \left(\frac{r}{\alpha}\left(e^{-\alpha t} - 1\right) + (E[0] - \epsilon[0])e^{-\alpha t}\right)^2. \quad (4)$$

## The profile of environmental change

For simplicity, let us consider that the CAS under study has been through a relatively stable environment, i.e., for $t \leq 0$, $\epsilon[t] = E[t]$, and, thus without loss of generality, we set $E[0] = \epsilon[0] = 0$. Then, we look at an environmental perturbation in the form of a ramp, characterized by a change of the external environment from its baseline value to a maximum value, $E_{max}$, at a fixed rate of change, $r$, phase I. Afterwards, the external environment remains constant at $E_{max}$, phase II. We write $E[t]$ as

$$E[t] = \{r t, \ t < t_{max} = E_{max}/r \ E_{max}, \ t \geq t_{max}. \quad (5)$$

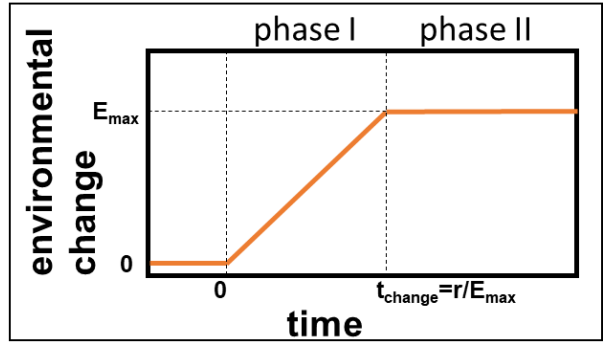

This corresponds to two distinct responses of the system,

$$\frac{dx}{dt} = -x^2 + 1 - \left(\frac{r}{\alpha}\left(e^{-\alpha t} - 1\right)\right)^2, \ t \leq t_{change} = E_{max}/r \quad (6.1)$$

$$\frac{dx}{dt} = -x^2 + 1 - \left(\frac{r}{\alpha}\left(e^{-\alpha t_{change}} - 1\right)e^{-\alpha(t - t_{change})}\right)^2, \ t > t_{change}. \quad (6.2)$$

## Characterizing the behavior of the system

### Phase I: environmental change

The initial phase is characterized by the rate of external environmental change, $r$, and it occurs while the environmental change is below $E_{max}$. In this initial phase, $b = 1 - \left(\frac{r}{\alpha}\left(e^{-\alpha t} - 1\right)\right)^2$, which controls the instantaneous basin of attraction and existence of the stable attractor, going from $b=1$ at $t=0$ to $b=1-(r/\alpha)^2$. **Figure S 1** shows the change in the basin of attraction for different rates of external environmental change relative to the adaptation rate. We see a reduction of the instantaneous basin of attraction of the prosperous states over time. If $\frac{r}{\alpha} < 1$, given enough time, the instantaneous basin of attraction converges to the region $\left(-\sqrt{1-\left(\frac{r}{\alpha}\right)^2}, \infty\right)$. For $\frac{r}{\alpha} > 1$, the instantaneous basin eventually disappears, and the CAS goes through a rate-induced transition (i.e., a transition from the prosperous state to the degraded state caused by the rate of change but not by the value of the external environment).

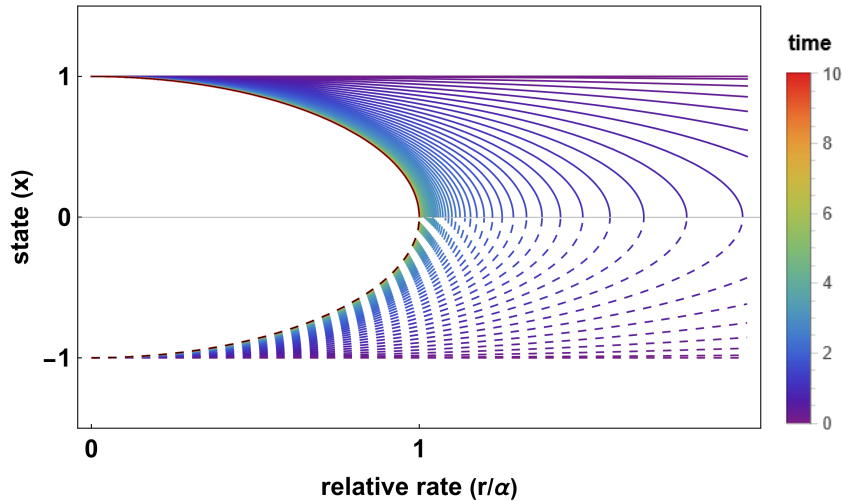

*Figure S 1. Changes in the position of the prosperous state (solid line) and lower bound of the basin of attraction (dashed line) for different rates. For a given rate of change (r) relative to the adaptation rate(α), r/α, the basin of attraction shrinks over time. For relative rates larger than 1 (r/α>1), there is a timepoint at which the basin of attraction disappears, and the system degrades towards -1 and beyond, given enough time. The exact moment the basin disappears depends on α, set to 1 in this illustration.*

**Figure S 2** illustrates the dynamics of $x$ over time, as well as the position of the attractor and the lower bound of its instantaneous basin of attraction. For low rates, r/α<1, the attractor always exists through time, even if the lower bound of its instantaneous basin of attraction shrinks. For high rates of change, there is a moment in which the attractor disappears.

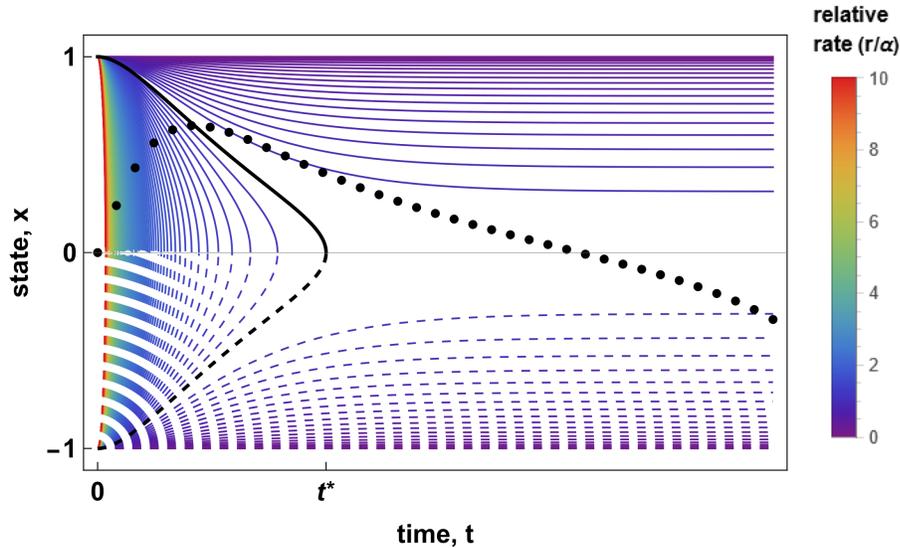

*Figure S 2. Changes in the position of the prosperous state (solid line) and lower bound of its basin of attraction (dashed line) over time. For low enough relative rates, r/α<1, the attractor (solid line) is sustained through time and the system can reach a prosperous equilibrium in the range (0,1) and tolerate arbitrarily large changes environmental changes; however, this is accompanied with a reduction of the basin of attraction (with a lower bound represented by the dashed line) and, thus, lower resilience to shocks. As the rate increases, there is a critical value of the environment above which the attractor disappears and no configuration of the system leads to recovery. The time evolution with r = 1.05 > α = 1 is illustrated by the dotted black line, which represents the value of x over time; its attractor corresponds to the black solid and the black dashed curve represents the position of the lower bound of the basin of attraction of the attractor.*

## Phase II: stabilized new environment

In the final phase, the initial state of $x$ (as well as $E$ and $\epsilon$) is simply the final value of the previous phase. Now, the process is reverted. Namely, if in the initial phase $b$ goes from 1 to a smaller value, $1 - \left(\frac{r}{\alpha}\left(e^{-\alpha t_{max}} - 1\right)\right)^2$, in this final phase, $b$ recovers back to 1. This means that **Figure S 1** (as well as the position of the attractor and its basin lower boundary in **Figure S 2**) still applies but with time in reverse. Thus, eventually, the system will either end up in a prosperous state normalized at 1 or tipped into a degraded state with $x < -1$.

# Regimes of Complex Adaptive Systems

It is trivial to see that if, at any point in time, $x$ degrades below -1, then it will always continue to degrade. It is also clear that if the environmental change is small (small $E_{max}$), the system will be able to track the prosperous equilibrium even if the change is instantaneous (as an example, take an initial condition of $x_0 = 1$ and any $E_{max} < 1$, or, more generally any point $x_0 > -\sqrt{1 - E_{max}^2}$). Thus, we can split the dynamics into three regions: 1) a "safe space" where the system will always track the prosperous equilibrium, as long as $E_{max}$ and perturbation to the equilibrium state are small, independently of $\frac{r}{\alpha}$; 2) a region where the system ends up (or starts) in the "irreversible tipping" region degraded beyond -1; and 3) a region where a combination of maximum environmental change and the rate to reach it are necessary to determine whether the system can track the prosperous equilibrium or whether it irreversibly tips. These three regimes are represented in Figure 1 in the main text.

## Coupled Complex Adaptive Systems

In the single CAS we introduced, the dynamics are determined by the structural adaptability, α, and the profile of external environmental change (r and Emax). Now, we want to understand the impact of having interacting CAS where they affect the structural adaptability or the external environment of others. Let us consider a set of $N$ coupled CAS. We take the equations for the state of each CAS $i$ as

$$\frac{dx_i}{dt} = -x_i^2 + 1 - (E_i - \epsilon_i)^2, \quad \frac{d\epsilon_i}{dt} = \alpha_i(E_i - \epsilon_i),$$

Where $\alpha_i$ can differ per CAS.

## Network contribution to structural adaptation

Consider the case all CAS share the same external environment $E_i = E$ but different CAS contribute the structural adaptation rate of other CAS, such that

$$\alpha_i = \alpha^* \sum_j A_{ij} \lfloor \frac{1+x_j}{2} \rfloor_0$$

Where $A_{ij}$ is an adjacency matrix which determines whether two nodes are connected and is normalized such that $\sum_j A_{ij} = 1$ for all CAS and $\lfloor \frac{1+x_j}{2} \rfloor_0 = \max(0, \frac{1+x_j}{2})$ is continuous and guarantees adjacent systems, j, only contribute to the adaptability of a focal system $i$ when their index is above -1 (i.e., before tipping).

# Additional results

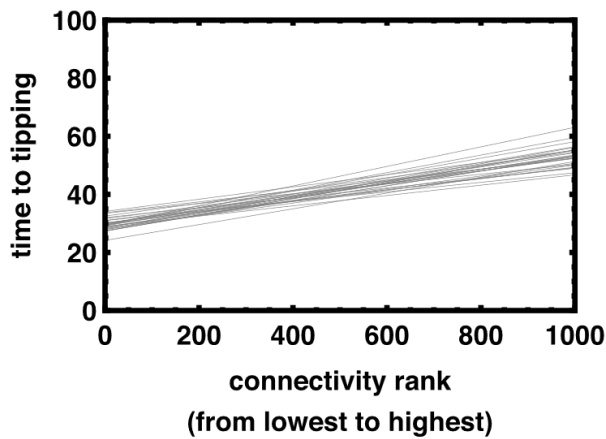

Figure S3. (Left) Role of degree in time to tipping. (Right) Role of centrality in time to tipping. Same setup as Figure 4, the right panel with r=0.475, in the critical transition phase.

Figure S4. The role of network topology. Same as Figure S3, for a random network.